\RequirePackage{fix-cm}
\documentclass[twocolumn,epjc3]{svjour3}  
\smartqed  
\usepackage[normalem]{ulem}  
\usepackage[dvipdfmx]{graphicx} 
\usepackage{color}
\renewcommand\sout[1]{\bgroup \color{red} \ULdepth=-.5ex \ULset {#1}}

\newcommand{\DDbx}{\{\!D\bar{D}^*\!\}} 
\newcommand{\DDbxn}{\{\!D^0\bar{D}^{*0}\!\}} 
\newcommand{\DDbxc}{\{\!D^{\!+}\!D^{*\!-}\!\}} 

\usepackage{amsmath,amssymb}
\usepackage{multirow}
\usepackage{longtable}
\usepackage{epstopdf}
\usepackage{times}
\usepackage[normalem]{ulem}  
\usepackage{bm}
\allowdisplaybreaks[4]
\journalname{Eur. Phys. J. A}
\begin{document}

\title{Femtoscopic study on $DD^*$ and $D\bar{D}^*$ interactions for  $T_{cc}$ and  $X(3872)$ \thanksref{t1}
}

\thankstext{t1}{Report No.: YITP-22-26}
\author{Yuki Kamiya\thanksref{e1,addr1,addr2}
        \and
        Tetsuo Hyodo\thanksref{e2,addr3,addr2} 
        \and
        Akira Ohnishi\thanksref{e3,addr4}
}

\thankstext{e1}{e-mail: kamiya@hiskp.uni-bonn.de}
\thankstext{e2}{e-mail: hyodo@tmu.ac.jp}
\thankstext{e3}{e-mail: ohnishi@yukawa.kyoto-u.ac.jp}


\institute{Helmholtz Institut f\"ur Strahlen- und Kernphysik and Bethe Center for Theoretical Physics, Universit\"at Bonn, D-53115 Bonn, Germany\label{addr1}
           \and
    RIKEN Interdisciplinary Theoretical and Mathematical Science Program (iTHEMS), Wako 351-0198, Japan\label{addr2}
           \and           
 Department of Physics, Tokyo Metropolitan University, Hachioji 192-0397, Japan\label{addr3}
 \and
 Yukawa Institute for Theoretical Physics, Kyoto University, Kyoto 606-8502, Japan\label{addr4}
}

\date{Received: date / Accepted: date}

\maketitle

\begin{abstract}
We investigate $DD^*$ and $D\bar{D}^*$ momentum correlations in high-energy collisions to elucidate the nature of $T_{cc}$ and $X(3872)$ exotic hadrons. 
Single range Gaussian potentials with the channel couplings to the isospin partners are constructed based on the empirical data.
The momentum correlation functions of the $D^0D^{*+}$, $D^+D^{*0}$, $D^0\bar{D}^{*0}$, and $D^+D^{*-}$ pairs are computed with including the coupled-channel effects.
We discuss how the nature of the exotic states are reflected in the behaviors of the correlation results. 


\end{abstract}

\section{Introduction}\label{sec:Intro}

The study of various exotic resonances in heavy quark sectors has been one of the most interesting subjects in recent hadron physics~\cite{Hosaka:2016pey,Guo:2017jvc,Brambilla:2019esw}.
The most extensively studied state is the $X(3872)$ lying just below the $D\bar{D}^*$ threshold, which is listed as $\chi_{c1}(3872)$ in the current PDG paper~\cite{ParticleDataGroup:2020ssz}. 
Ever since its first observation in 2003~\cite{Choi:2003ue}, this exotic hadron has attracted huge interest of researchers and a bunch of the experimental and theoretical studies have been devoted to understand this state.
Nevertheless, its nature still remains to be elucidated. 

Recently, the LHCb Collaboration reported a clear signal of the doubly charmed tetraquark state $T_{cc}^+$ in the mass spectrum of $D^0D^0\pi^+$~\cite{LHCb:2021vvq,LHCb:2021auc}.
Such exotic states with two heavy quarks  and two light antiquarks are theoretically predicted with the quark model in Refs.~\cite{Ballot:1983iv,Zouzou:1986qh} more than thirty years ago. 
In contrast to the $X(3872)$, this $T_{cc}$ state is found in the genuine exotic channel, which requires at least four valence quark components ($cc\bar{u}\bar{d}$).
Although the $X(3872)$ and $T_{cc}$ are in different sectors, there is one similarity between them, {\it i.e.}, the existence of a nearby two-meson threshold. Namely, the $T_{cc}$ peak is also found just below the $DD^*$ threshold. The proximity with the $D\bar{D}^*$ and $DD^*$ thresholds would imply the molecular nature of these states. 
It should be noted, however, that the structure of X(3872) is still under debate.
In the study of Ref.~\cite{Bignamini:2009sk}, it is shown that the contribution of the $c\bar{c}$ component is important 
by the analysis of the prompt production cross section.
On the other hand, the enhancement of the production yield in $AA$ collisions observed in CMS~\cite{CMS:2021znk}
seems to imply that $X(3872)$ contains a significant fraction of  the hadronic molecule component~\cite{ExHIC:2017smd}.
In order to discriminate the possible structures of $X(3872)$, it is desirable to experimentally access the $D\bar{D}^*$ interaction.

For the study of the near-threshold resonances, the femtoscopy using the two-particle momentum correlation function in high-energy  collisions is a helpful technique because the correlation function is sensitive to the low-energy hadron interactions.
With the femtoscopy, various interactions in the strangeness sector have been investigated theoretically~\cite{Morita:2014kza
,Ohnishi:2016elb
,Morita:2016auo
,Hatsuda:2017uxk
,Mihaylov:2018rva
,Haidenbauer:2018jvl
,Morita:2019rph
,Kamiya:2019uiw
,Kamiya:2021hdb} 
and experimentally~\cite{STAR:2014dcy
,ALICE:2017jto
,STAR:2018uho
,ALICE:2018ysd
,ALICE:2019hdt
,ALICE:2019buq
,ALICE:2019eol
,ALICE:2019gcn
,ALICE:2020mfd
,ALICE:2021szj
,ALICE:2021cpv
,Fabbietti:2020bfg}. 
It turns out that the source size dependence of the correlation function is useful to distinguish the existence or non-existence of hadronic bound states~\cite{Morita:2019rph,Kamiya:2019uiw}.
Recently, the $D^{-}p$ correlation function has been measured by the ALICE collaboration~\cite{ALICE:2022enj}, which paves the way to the femtoscopy in the charm sector.

In this study, we discuss the correlation functions of the $DD^*$ and $D\bar{D}^*$ channels towards the understanding of the nature of the  $T_{cc}$ and $X(3872)$ states. To this end, we construct one-range Gaussian potentials for the $DD^*$ and $D\bar{D}^*$ channels which reproduce the empirical information in these channels. Including the coupled-channel effects with the isospin partners and the decay channels, we compute the correlation functions.

This paper is organized as follows.
In Sect.~\ref{sec:Method}, we construct the $DD^*$ and $D\bar{D}^*$ potentials from the empirical data and summarize the method to calculate the correlation function with coupled-channel source effect. In Sect.~\ref{sec:results}, we show the results of the correlation functions of the $D^0D^{*+}$, $D^+D^{*0}$, $D^0\bar{D}^{*0}$, and $D^+D^{*-}$ channels and discuss how the exotic states can be studied in the future femtoscopy experiments. 
Section~\ref{sec:summary} is devoted to summarize this study.

\section{Method}\label{sec:Method}

\begin{table*}
		\caption{Strength parameters $V_0$ for the $DD^*$ and $D\bar{D}^*$ potentials and the scattering lengths in the $DD^*$ and $D\bar{D}^*$ channels. The scattering lengths of the lower channels (third column) are the empirical inputs.  }
		\label{tab:V0}
	\begin{center}
		\begin{tabular}{cccc}
			\hline\hline
			$DD^*$& $V_0$ [MeV] &  $a_0^{D^0D^{*+}}$ [fm] & $a_0^{D^+D^{*0}}$  [fm] \\
			\hline
			&$-36.569-i1.243$&$-7.16+i1.85$ & $-1.75+i1.82$ \\
			\hline
			$\DDbx$&$ V_0$ [MeV] &  $a_0^{\DDbxn}$ [fm] & $a_0^{\DDbxc}$  [fm] \\\hline
			& $-43.265-i6.091$ &$-4.23+i3.95$ & $-0.41 +i1.47$  \\
			\hline\hline
		\end{tabular}
	\end{center}
\end{table*}

Let us first summarize the relevant channels which couple to the system of interest. For the $T_{cc}$ and $X(3872)$ states, we cannot neglect the mass difference among the isospin multiplets, because the deviation of the eigenenergy from the threshold is comparable or smaller than the isospin breaking effect. The $T_{cc}$ locates just below the $D^{0}D^{*+}$ threshold, and it also couples to the $D^{+}D^{*0}$ channel whose threshold lies slightly above that of the $D^{0}D^{*+}$ channel. At energies lower than the $T_{cc}$, the three-body $DD\pi$ channels are open, which provide the finite decay width of $T_{cc}$. The $X(3872)$ lies just below the $\DDbxn=(D^{0}\bar{D}^{*0}+\bar{D}^{0}D^{*0})/\sqrt{2}\ (C=+)$ threshold and couples also to the higher energy $\DDbxc=(D^{+}D^{*-}+D^{-}D^{*+})/\sqrt{2}\ (C=+)$ channel. At much lower energies, the decay channels such as $\pi\pi J/\psi$ couple to the $X(3872)$. In the following, we explicitly treat the $D^{0}D^{*+}$ and $D^{+}D^{*0}$ channels for $T_{cc}$ and $\DDbxn$ and $\DDbxc$ channels for $X(3872)$, and the decay effect to the other channels are renormalized in the imaginary part of the potential. Thus, the Hamiltonian of the system is expressed by a $2\times 2$ matrix in the channel basis. 

Next, we construct the $DD^*$ and $D\bar{D}^*$ potentials. Assuming that the interaction is isospin symmetric, the strong interaction part of the coupled-channel potentials can be given by the $I=0$ and $I=1$ components as 
\begin{align}
	V_{DD^*\text{/}D\bar{D}^*} =\frac{1}{2} \left(
	\begin{array}{cc}
		V_{I=1}+V_{I=0}~  & ~V_{I=1}-V_{I=0} \\
		V_{I=1}-V_{I=0}~  & ~V_{I=1}+V_{I=0}
	\end{array}
	\right) ,
	\label{eq:Vstrong}
\end{align}
where we assign channel $i=1$ and 2 to $D^+D^{*0}$ and $D^0D^{*+}$ for the $DD^*$ system and $\DDbxn$ and $\DDbxc$ for the $D\bar{D}^*$ system, respectively. Because the $T_{cc}$ and $X(3872)$ couples to the $I=0$ channel, we assume that the $I=0$ component gives the dominant contribution, and set 
\begin{align}
	V_{I = 0}&= V(r),\\
	V_{I = 1}&= 0,
\end{align}
where $V(r)$ is a spherical Gaussian potential:
\begin{align}
	V(r) = V_0 \exp(-m^2r^2),
\end{align}
where $V_0$ is the interaction strength and $m$ is the parameter of the dimension of mass to control the range of the interaction.
Here we use the charged (isospin averaged)  pion mass $m_{\pi^\pm}$  ($m_\pi$) for the  $DD^*$ ($D\bar{D}^*$) interactions because the lightest exchangeable meson, pion, determines the interaction range. Thus, in this formulation, we are left with a single parameter $V_{0}$ for each $DD^*\text{/}D\bar{D}^*$ potential. Note that $V_{0}$ takes a complex number, in order to express the decay effects into the lower energy channels. While the $DD^{*}$ potential is free from the Coulomb interaction, for the $\DDbxc$ channel we should include the Coulomb force:
\begin{align}
	V_{D\bar{D}^*}^{c}(r) = \left(
	\begin{array}{cc}
		0  & 0 \\
		0  & -\alpha/r
	\end{array}
	\right) ,
\end{align}
with the fine structure constant $\alpha$. This potential is added to Eq.~\eqref{eq:Vstrong} for the $D\bar{D}^*$ potential.

Here we determine the potential strength $V_0$ so as to reproduce the empirical data for these systems. For the $DD^{*}$ potential, we use the scattering length $a_0^{D^0D^{*+}}=-7.16+i1.85$ fm, given in the experimental analysis in Ref.~\cite{LHCb:2021auc}.\footnote{Here we use the the high-energy physics convention for the scattering length where the positive (negative) real value corresponds to the weakly attractive (repulsive or strongly attractive) interaction.} For the $D\bar{D}^{*}$ potential, we use the scattering length $a_0^{\DDbxn}=-4.23+i3.95$ fm which is determined by the eigenenergy $E_{h}=-0.04-i0.60$ MeV in PDG~\cite{ParticleDataGroup:2020ssz} measured from the $D^0\bar{D}^{*0}$ threshold, as $a_{0}^{\DDbxn}=-i/\sqrt{2\mu E_{h}}$ with the reduced mass $\mu$. We notice that these scattering lengths have a much larger magnitude than the typical length scale of the strong interaction $\sim 1$ fm. The obtained potential strengths are summarized in Table~\ref{tab:V0}.
For the later use, the scattering lengths of the higher channels ($D^+D^{*0}$ and $\DDbxc$) calculated with the same potentials are also listed. 
Note that all these calculations are performed in the coupled-channel scheme. 

\begin{figure}
	\begin{center}
		\includegraphics[width=0.48\textwidth]{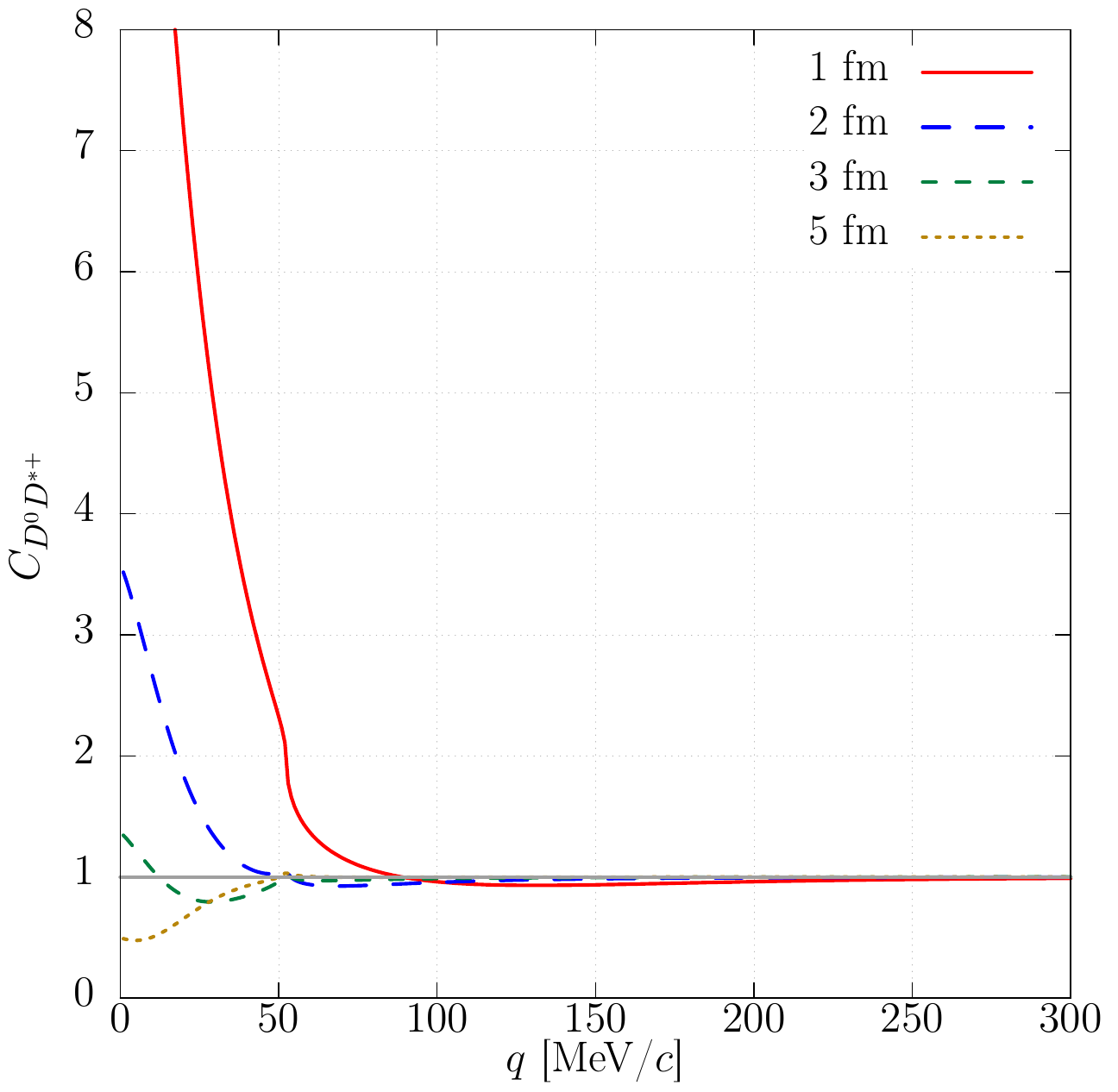}
		\includegraphics[width=0.48\textwidth]{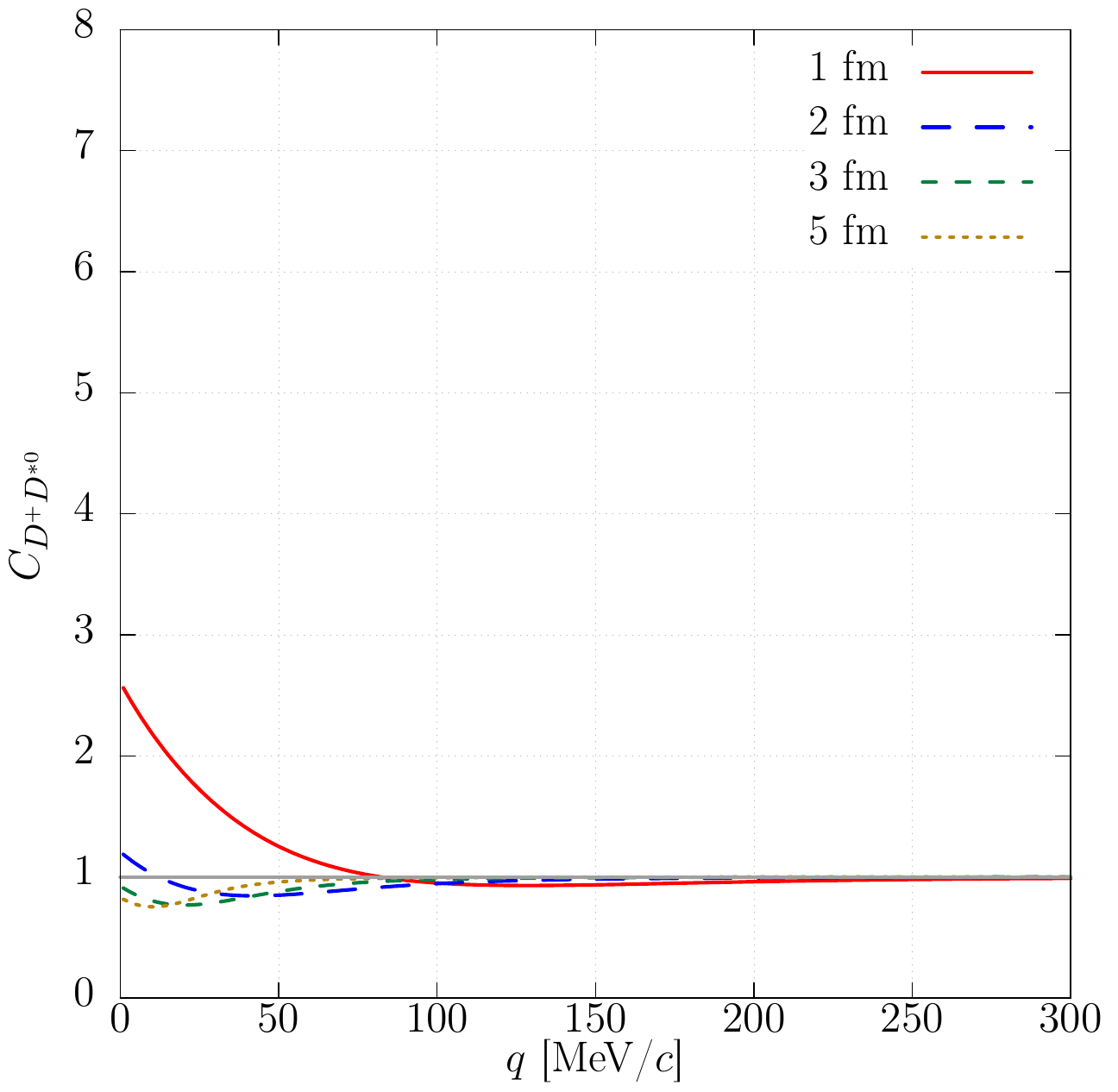}
		\caption{The correlation functions of the $D^0D^{*+}$ (top) and $D^+D^{*0}$ (bottom) pair with the source size $R=1,2,3$, and $5$ fm.}	
		\label{fig:corr_DD*}
	\end{center}	
\end{figure}

To calculate the correlation functions $C(q)$ with the coupled-channel effects, we employ the Koonin-Pratt-Lednicky-Lyuboshitz-Lyuboshitz formula (KPLLL) formula~\cite{Lednicky:1998r,Haidenbauer:2018jvl,Kamiya:2021hdb} given by
\begin{equation}
	C(q)= \int d^3r \sum_{i=1}^{2} \omega_{i} S_i(r) |\Psi^{(-)}_{i}(q;\bm{r})|^2
	\ ,\label{Eq:KPLLL}
\end{equation}
where the wave function $\Psi_i^{(-)}$ in the $i$-th channel is written as a function of the relative coordinate $\bm{r}$, with imposing the outgoing boundary condition on the measured channel. 
We consider the small momentum region and assume that only the $s$-wave component of the wave function $\Psi_i^{(-)}$  is modified by the strong interaction. 
The wave function is calculated by solving the Schr\"odinger equation with the hermite conjugated potential $V^\dagger$, which gives the appropriate boundary condition for the eliminated decay channels (See ~\ref{sec:Vstar}). 
We adopt a common static Gaussian source function for all the channels $S_i(r) =\exp(-r^2/4R^2)/(4\pi R^2)^{3/2}$ with the source size $R$, and the weight factor $\omega_i$ is taken as unity for all channels. 
The weight factor $\omega_i$ represents the ratio of the pair production yield in the $i$th channel with respect to the measured channel. Since
 we only include the coupled-channel effect of the isospin partners, they are considered to have the equivalent emitting source. The source size $R$ ranges from $\sim 1$ fm for the high-multiplicity events in $pp$ collisions to $\sim (5\!-\!6)$ fm for the central PbPb collisions.


While we construct the $D\bar{D}^*$ potential in the charge conjugation $C\!=\!+$ combination which couples to the $X(3872)$, the experimental measurement of the correlation function will be done with fixed charge states, {\it i.e.}, either $D^0\bar{D}^{*0}$ or $\bar{D}^0D^{*0}$. To obtain the correlation functions of the fixed charge states, the correlation functions in the $C\!=\!-$ sector are also needed to take an average of the $C\!=\!+$ and $C\!=\!-$ contributions. In this exploratory study, we assume that the $C\!=\!-$ interaction is small and can be neglected with respect to the dominant $C\!=\!+$ contribution. In this case, we obtain the experimentally accessible correlation functions from the correlation function calculated by the $C\!=\!+$ potential as 
\begin{align}
C_{D^0\bar{D}^{*0}}&	
=C_{\bar{D}^0D^{*0}}
=\frac{1}{2}\left(C_{\DDbxn}+1\right) ,	\label{C_D0Dbars0} \\
C_{D^{\!+}D^{*\!-}}&
=C_{D^{\!-}D^{*\!+}}
=\frac{1}{2}\left(C_{\DDbxc}+C_{\mathrm{pure \ Coul.}}\right),\label{C_DmDsm}
\end{align}
where $C_{\mathrm{pure \ Coul.}}$ is calculated only with the Coulomb interaction by switching off the strong interaction contribution. 

\section{Results}\label{sec:results}

Now we calculate the correlation functions with the constructed potentials.
First we show the $DD^{*}$ sector coupled with the $T_{cc}$ state.
The correlation function of the $D^0D^{*+}$ and the $D^+D^{*0}$ pairs with source sizes $R=1,2,3$, and $5$ fm are shown in Fig.~\ref{fig:corr_DD*}.
We can see that the source size dependence typical to the system with a shallow bound state for both correlation functions; the enhancement in the small source case turns to the suppression for the large source case~\cite{Kamiya:2021hdb}.
The stronger correlation is found in the $D^0D^{*+}$ channel,
whose threshold is closer to the $T_{cc}$ pole. 
The cusp structure is seen at the $D^+D^{*0}$ threshold ($q\simeq 52$ MeV/$c$) in the $D^0D^{*+}$ correlation, while the strength is not very prominent.

\begin{figure}
	\begin{center}
		\includegraphics[width=0.48\textwidth]{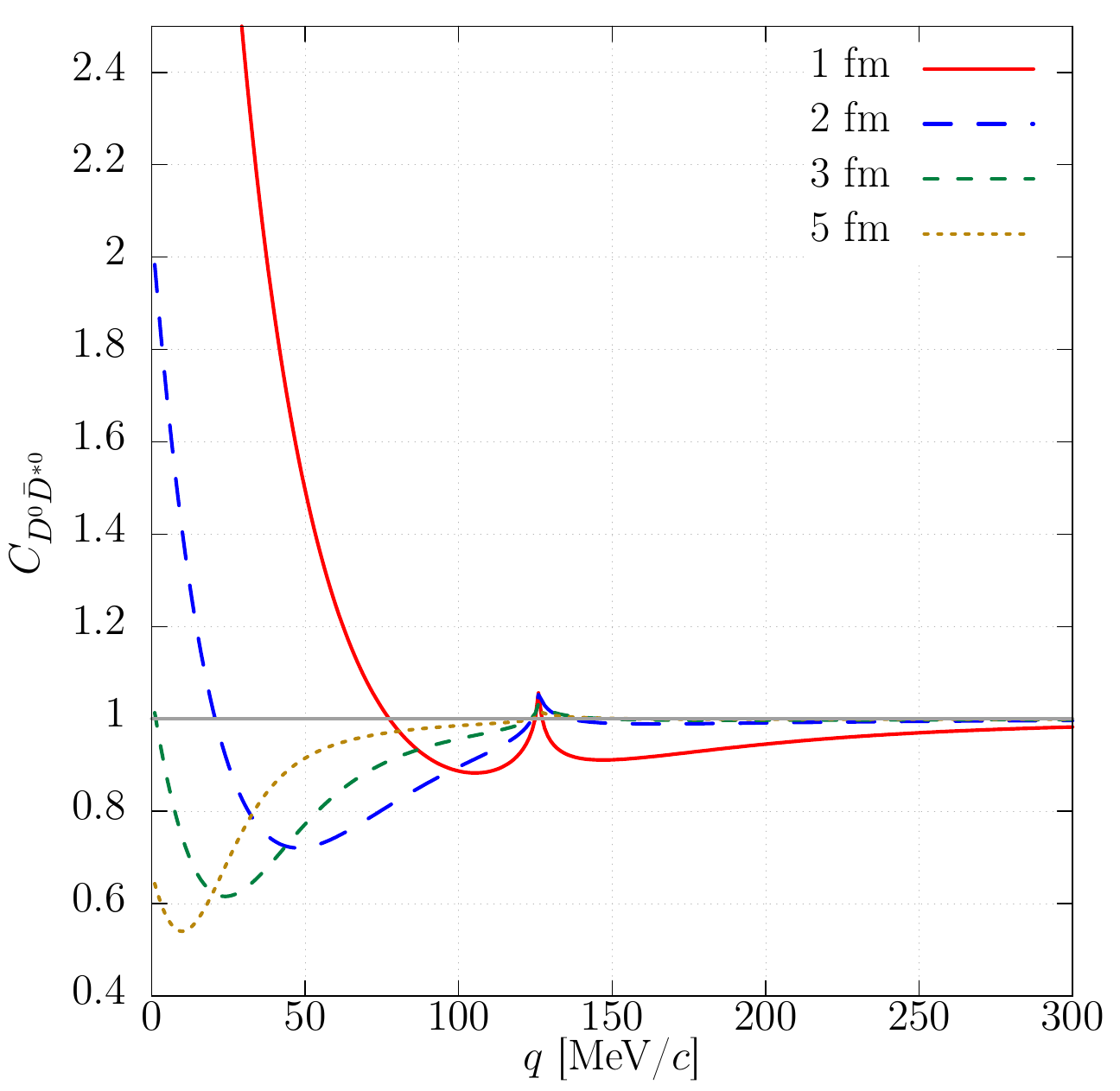}
		\includegraphics[width=0.48\textwidth]{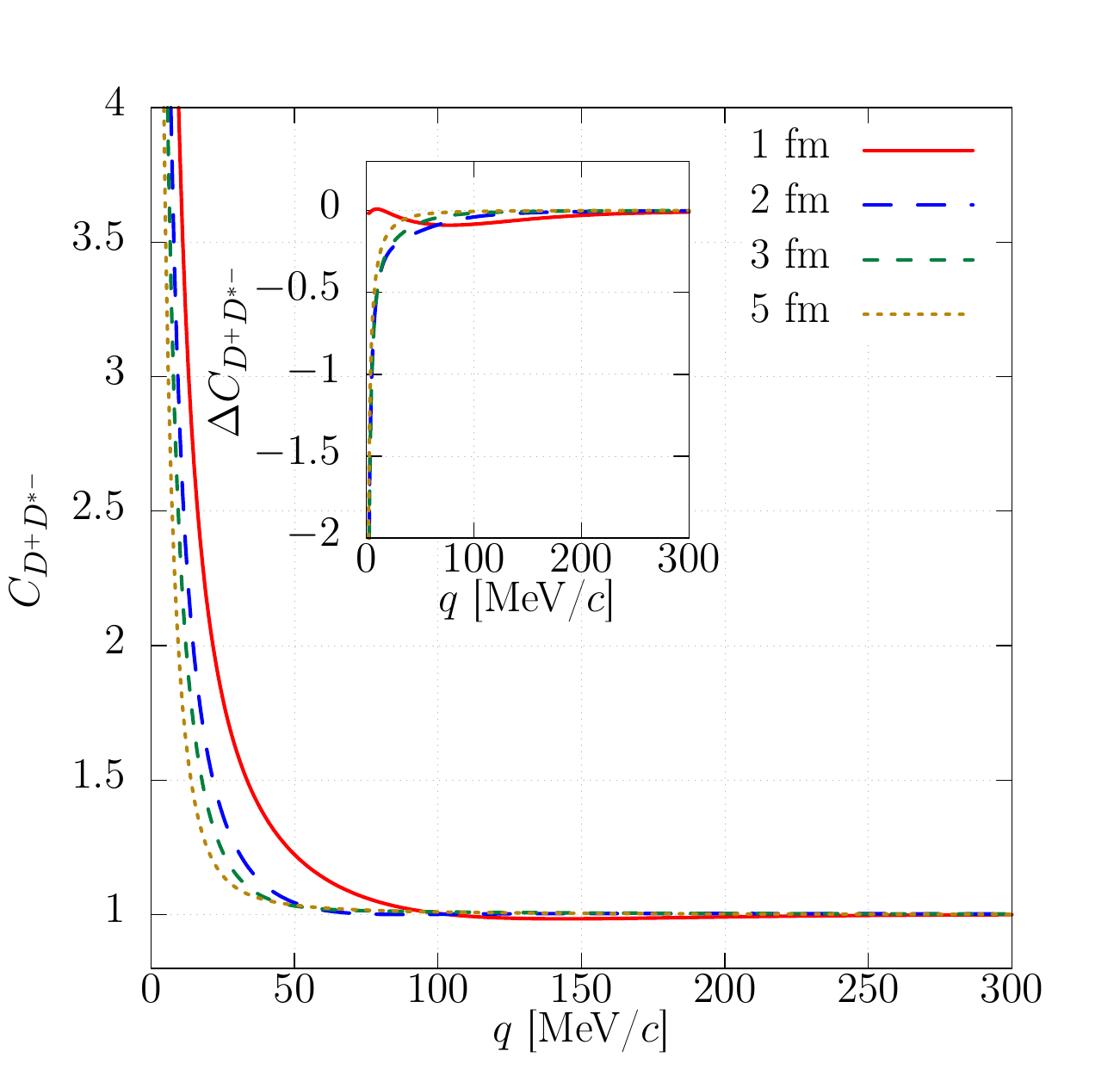}
		\caption{The correlation functions of the $D^0\bar{D}^{*0}$ (top) and $D^+D^{*-}$ (bottom) pair with the source size $R=1,2,3$, and $5$ fm.
		For $D^+D^{*-}$ pair, the difference from the pure Coulomb case $\Delta C $ is shown in sub figure. }	
		\label{fig:corr_DDbar}
	\end{center}		
\end{figure}

Next we show the results of the $D\bar{D}^{*}$ correlation function
coupled with the $X(3872)$ in Fig.~\ref{fig:corr_DDbar}. 
Here we plot the correlation functions of the fixed charges states in Eqs.~ \eqref{C_D0Dbars0} and \eqref{C_DmDsm} which can be compared with the experimental measurements.
The characteristic strong source size dependence with the shallow bound state is found in $C_{D^0\bar{D}^{*0}}$.
We can also see the cusp structure at the $D^+D^{*-}$ threshold ($q\simeq 126$ MeV/$c$).
The cusp structure is more prominent for the smaller source case. 
This is because the coupled-channel source effect by the ${D^+D^{*-}}$ channel is stronger for the smaller source case~\cite{Kamiya:2019uiw}.
On the other hand, due to the attractive Coulomb force, the $C_{D^+D^{*-}}$ correlations show a strong enhancement at small $q$. To extract the contribution by the strong interaction, we show the difference from the pure Coulomb case $\Delta C =C_{D^+D^{*-}} - C_{\mathrm{pure \ Coul.}}$. We can see that the effect of the strong interaction emerges mainly as the suppression compared to the pure Coulomb case. However, the deviation $|\Delta C|$ is less than 0.2 for the momentum region $q  >  50$ MeV/$c$.
Thus, the correlation of $D^+D^{*-}$ pair is expected to be dominated by the Coulomb contribution.

In this study, we used the empirically determined scattering lengths as input to calculate the correlation functions. 
Given the correlation data obtained from the precise future measurement, 
we can independently determine the scattering lengths $a_0$
because the correlation functions are sensitive to the low-energy interaction. 
According to the Weinberg's weak-binding relation~\cite{Weinberg:1965zz,Kamiya:2015aea,Kamiya:2016oao}, 
the compositeness, which is defined as the probability of finding molecular state in the eigenstate, is directly related to the ratio of the $a_0/R_h$ where $R_h$ is the length scale determined with the eigenenergy $E_h$ as $R_h=1/\sqrt{-2\mu E_h}$.
Thus, combined with the information of the pole position, 
to measure the these correlation functions leads to understand the nature of $T_{cc}$ and $X(3872)$ states. 
 

\section{Summary}\label{sec:summary}

We have studied the correlation functions of the $DD^*$ and $D\bar{D}^*$ pairs for the purpose of the investigation of the $T_{cc}$ and $X(3872)$ exotic states. 
With the assumption of the molecular nature of these states, one-range Gaussian potentials are constructed for the $DD^*$ and $D\bar{D}^*$ channels from the empirical data. 
Due to the large scattering lengths, 
the calculated correlation functions in the lower channels ($D^0D^{*+}$ and $D^0\bar{D}^{*0}$), which are closer to the exotic states, show the characteristic behavior of the bound state below the threshold.
On the other hand, the correlation function of the $D^+D^{*0}$ channel shows less prominent behavior due to the energy difference from the $T_{cc}$ pole, and the correlation in the $D^+D^{*-}$ channel is mainly caused by the Coulomb interaction. Given the successful measurement of the $D^{-}p$ correlation function by the ALICE collaboration~\cite{ALICE:2022enj}, we expect that the measurements of the $DD^*$ and $D\bar{D}^*$ correlations in future will bring new insights of the exotic hadrons from the viewpoint of the femtoscopy.

In this study, we have introduced the potentials in the channels that couple to the exotic states (isospin $I=0$ and charge conjugation $C=+$), and have neglected the interactions in the other channels. This is because the existence of near-threshold states implies the strong interaction, which is considered to give the dominant contribution for the correlation function. For more quantitative discussion of the correlation functions, these subleading effects should also be considered. In particular, the cusp structure may be sensitive to the isospin $I=1$ interaction, because the coupling between the isospin partners are given by the difference of the two isospin components. The $D\bar{D}^*$ interaction in the $C=-$ sector is still unclear at this moment, but the neutral partner of $Z_c(3900)$~\cite{BESIII:2015cld} may play an important role in this channel. These effect should be discussed in the future studies.

\appendix

\section{Outgoing boundary condition for the optical complex potential}\label{sec:Vstar}
The wave function for the KPLLL formula~\eqref{Eq:KPLLL} must satisfy the outgoing boundary condition where the flux of the outgoing wave of the reference channel is normalized to be unity.
On the other hand,  the complex optical potentials are constructed based on the scattering problem with the incoming boundary condition where the flux of the incoming wave is normalized.
This boundary condition is applied 
to the integrated channels, whose coupling to referenced 
channels  ($D^+D^{*0}$ and  $D^0D^{*+}$ in the case of $DD^*$ sector)
give the imaginary part of the potential. 
Thus, we cannot obtain the correct wave function $\psi$ by solving the Schr\"odinger equation 
\begin{align}
	H\psi=[H_0+V]\psi=E\psi \label{eq:Schrodinger},
\end{align}with the boundary condition only with the  referenced channels.

We claim that
we can just take the hermite conjugate of the potential $V$ and solve the Schr\"odinger equation in order to obtain the wave function which satisfies the boundary conditions for all the channels, 
\begin{align}
[H_0+V^\dagger ]\psi=E\psi\label{eq:Schrodinger_T},
\end{align}
with the outgoing boundary condition.
One can easily check that $\psi^*$ satisfies the original Schr\"odinger equation with incoming boundary condition. 

Taking the hermite conjugate of the potential $V$ corresponds to consider the time reversal of the system. 
This can be understood as follows.  
Let us consider the two channel scattering problem with spinless particles where channel 1 (2) has higher threshold energy and is measured (lower threshold energy and is not measured).
The Hamiltonian for this system is given as 
\begin{align}
H = & \left(\begin{array}{cc}
	H_{11}&H_{12}  \\
	H_{21}& H_{22}
\end{array}\right) = 
 \left(\begin{array}{cc}
	H_{11}^0 + V_{11}&V_{12}  \\
	V_{21}& H_{22}^0 +V_{22}
\end{array}\right) ,
\end{align}
where $H_{ij}^{0}$ and $V_{ij}$ are the free Hamiltonian and the interaction potential, respectively. 
The Lippmann-Schwinger equation for the $T$ matrix is given  as 
\begin{gather}
	T = V+ V G^0T,\\
	G^0 = \mathrm{diag.} (G^0_1, G^0_2),
\end{gather}
with the free propagator
\begin{align}
	G_i^0 (z) = (z -H_{ii}^0)^{-1}.
\end{align}
With the Feshbach projection~\cite{Feshbach:1958nx,Feshbach:1962ut} for channel 2, the Lippmann-Schwinger equation for channel 1 can be written with the effective potential $V_{\mathrm{eff}}$ as 
\begin{align}
T_{11}(z) &= V_{\mathrm{eff}}(z)+V_{\mathrm{eff}}(z)G^0_1(z) T_{11}(z), \\ 
V_{\mathrm{eff}}(z)& =V_{11} +V_{12}G_2(z)V_{21}, 
\end{align}
where $G_i(z)$ is the full propagator given as 
\begin{align}
G_i (z) = (z -H_{ii})^{-1}.
\end{align}
The contour of the time integration of $G_i^{(0)}(z)$ can be chosen by taking $z\rightarrow  E + i\epsilon$ for the scattering problem. 
On the other hand, that of the time-reversed system can be given as  $z\rightarrow  E - i\epsilon$. 
This effective potential is complex due to the pole term included in $G_2(E-i\epsilon)$.
Then the effective potential in the time reversed system is given as 
\begin{align}
	V_{\mathrm{eff}} (E-i\epsilon)& = V_{11} + V_{12}G_2(E-i\epsilon)V_{21} \notag\\ 
	                                   & = V_{11}^\dagger + V_{21}^\dagger G_2^\dagger(E+i\epsilon)V_{12}^\dagger \notag \\
	                                   & = \left[V_{11} + V_{12}G_2(E+i\epsilon)V_{21}  \right]^\dagger \notag\\ 
	                                   & = V_{\mathrm{eff}}^\dagger(E+i\epsilon).
\end{align}
Here we assumed that the full Hamiltonian is hermitian and the potential $V$ is real. 
Thus, the hermite conjugated effective potential corresponds to that in the time reversed system.
Remembering that the time reversal operator $T$ acts on the wave function as $T\psi= \psi^*$~\cite{Taylor},
the system obtained from Eq.~\eqref{eq:Schrodinger_T} with the outgoing boundary condition corresponds to the time-reversed system written with Eq.~\eqref{eq:Schrodinger} with incoming boundary condition.

The imaginary part of the optical potential causes the suppression or the enhancement of the wave function component of the referenced channel depending on its sign.
In the scattering problem of the coupled-channel system, 
the asymptotic form of the $s$-wave component of the scattering wave function of channel 1 is given with the $S$ matrix  component as 
\begin{align}
	\psi_1(q;r) \rightarrow \frac{1}{2iqr}\left( e^{-iqr} - \mathcal{S}_{11}  e^{iqr} \right). \label{eq:BC_incoming}
\end{align}
Due to the coupling to  channel 2, the absolute value of the $S$ matrix component $S_{11}$ is less than unity, which leads the reduced outgoing wave ($e^{iqr}$) compared to the normalized incoming wave ($e^{-iqr}$). 
When we use the complex optical potential $V$ with negative imaginary part, 
this reduction of the wave function is caused by the imaginary part of the potential. 
On the other hand, 
the outgoing boundary condition, which is used for the correlation study, is given as 
\begin{align}
	\psi_1(q;r) \rightarrow \frac{1}{2iqr}\left( e^{iqr} - \mathcal{S}^{\dagger}_{11}  e^{-iqr} \right).\label{eq:BC_outgoing}
\end{align}
In this case, the flux of the outgoing wave (1) is larger than that of the incoming wave ($|\mathcal{S}_{11}^\dagger|$). 
This is because the wave function of  channel 2 flows into channel 1 by the coupling potential to give the normalized outgoing wave. 
When we use the hermite conjugated optical potential $V^\dagger$ with positive imaginary part, 
its imaginary part  causes the enhancement of the channel 1 component. 
We also note that the resulting wave function can also be obtained
by solving Eq.~\eqref{eq:Schrodinger} with the (standard) incoming boundary condition
and taking the complex conjugate of the wave function.

%

\begin{acknowledgements}
The authors thank Laura Fabbietti and Fabrizio Grosa for useful discussions.
This work has been supported in part by the Grants-in-Aid for Scientific Research
from JSPS (Grant numbers
JP21H00121, 
JP19H05150, 
JP19H05151, 
JP19H01898, 
JP18H05402, 
and
JP16K17694), 
by the Yukawa International Program for Quark-hadron Sciences (YIPQS),
by the Deutsche Forschungsgemeinschaft (DFG) and the National Natural Science Foundation of China (NSFC) through the funds provided to the Sino-German Collaborative Research Center ``Symmetries and the Emergence of Structure in QCD" (NSFC Grant No. 12070131001, DFG Project-ID 196253076 -- TRR 110).

\end{acknowledgements}



\end{document}